\documentclass[conference]{IEEEtran}
\IEEEoverridecommandlockouts

\usepackage{cite}
\usepackage{amsmath,amssymb,amsfonts}
\usepackage{algorithmic}
\usepackage{graphicx}
\usepackage{textcomp}
\usepackage{xcolor}
\usepackage[hyphens]{url}
\usepackage{hyperref}
\hypersetup{colorlinks=true,allcolors=black}

\newcommand{\repo}{https://github.com/FudanSELab/train-ticket/blob/813dd01/}
\newcommand{\src}[2]{\href{\repo#1}{\nolinkurl{#2}}}
\usepackage{booktabs}
\usepackage{tabularx}
\usepackage{array}

\begin{document}

\title{CC4M: Code Clone Analysis and Visualization for Microservices}

\author{
\IEEEauthorblockN{Gen Kawamata, Yuki Ota, Norihiro Yoshida, Shiyu Yang, Erina Makihara, Katsuro Inoue}
\IEEEauthorblockA{
Ritsumeikan University\\
Osaka, Japan\\
\{is0665sp, is0607ih\}@ed.ritsumei.ac.jp,
\{norihiro, yangsy, makihara, inoue-k\}@fc.ritsumei.ac.jp
}
}

\maketitle

\begin{abstract}
Microservice architecture supports software evolution by 
decomposing a system into small, loosely coupled services 
that can be deployed independently.
Contrary to the expectation of high modularity, prior studies have reported that code clones exist across service boundaries, some of which are co-modified in the same version.
Such clones may require changes to be propagated across service boundaries, 
thereby undermining service independence and increasing maintenance costs.
However, existing tools do not support microservice-aware clone analysis.
We present CC4M, a microservice-aware clone analysis and visualization
tool. 
CC4M detects and enriches clone pairs with service-boundary, co-modification, file-category, and metric information. The enriched clones are visualized in an interactive scatter plot with explicit service boundaries, supporting metric-based filtering to prioritize clones with potentially higher maintenance impact.
Using an open-source microservice application, we illustrate how CC4M
helps identify the potential impact scope of code changes. 
A demo video and the tool are available at https://www.youtube.com/watch?v=0xOIQPFbkUg and https://doi.org/10.5281/zenodo.21204195, respectively.

% Moの研究では，same versionとの記述
% 2/3くらいまで減らす（背景を減らしてもいいかも）
% file カテゴリに統一

% マイクロサービスアーキテクチャは，システムを小規模で疎結合かつ
% 独立してデプロイ可能なサービス群へ分割することで，ソフトウェア進化を支援する．
% しかし，先行研究では，マイクロサービスシステムにおいて
% サービス間コードクローンが存在し，その一部が同一コミット内で同時修正されていることが
% 報告されている．このようなクローンは，サービス境界を越えて変更を波及させる可能性があり，
% サービスの独立性を損ない，保守コストを増加させる要因となる．
% 本稿では, マイクロサービスシステムを対象としたコードクローン検出, 可視化ツール
% である CC4M を提案する. CC4M は Type-2 クローンペアを検出し,
% それらにサービス境界, 同時修正, およびファイルカテゴリの情報を付与した上で,
% service span や co-modification frequency などのクローンメトリクスを計算する.
% 拡張されたクローンは, 明示的なサービス境界を持つインタラクティブな散布図として
% 可視化され, CC4M はメトリクスに基づくフィルタリングにより, 保守上の影響が
% 大きいクローンの優先付けを支援する.
% あるオープンソースのマイクロサービスアプリケーションを用いて, コード変更の潜在的な
% 影響範囲の特定やリファクタリング候補の発見など, CC4M がソフトウェア進化タスクを
% どのように支援するかを示す.
% デモ動画は YouTube で, ツールは Zenodo (DOI) で公開する. GitHub へは Zenodo と本文脚注から辿れる.

\end{abstract}


\section{Introduction}

Microservice architecture decomposes a software system into small, loosely
coupled, and independently deployable services. It has been widely adopted
in open-source and industrial projects to support software evolution
through modularity and technology heterogeneity
\cite{Richardson2018MicroservicesPatterns,Newman2021BuildingMicroservices,LewisFowler2014Microservices}.
% 企業名を消去
%It has been widely adopted
%in both open-source and industrial projects, including systems at Netflix,
%Amazon, and Spotify, to support software evolution through modularity and
%technology heterogeneity
%\cite{Richardson2018MicroservicesPatterns,Newman2021BuildingMicroservices,LewisFowler2014Microservices}.
% In a microservice system, each service may be implemented using different
% technologies and is expected to evolve independently.

Because microservices emphasize modularity and service independence, one might expect clones across service boundaries to be limited. 
% Code clones
% are code fragments that are identical or similar to one
% another~\cite{RoyCordy2009}. 
% Such pairs and sets of similar fragments are called clone pairs and clone sets, respectively.
However, Mo et al. reported that code clones exist across
microservices in open-source microservice projects, and showed that
clones are co-modified across service boundaries between consecutive releases~\cite{Mo2021ESEM}.
% 追加
Here, code clones are identical or similar code fragments~\cite{RoyCordy2009}; 
a clone pair is a pair of such fragments, 
and a clone set is a set of mutually similar fragments.
Co-modification refers to the case in which the code fragments
constituting a clone pair are modified together in the project's revision history.
% versionの表現を修正
Following prior work~\cite{Mo2021ESEM}, we classify clone
pairs in microservice systems based on service boundaries. 
A clone pair is called a \emph{within-service clone pair} when both fragments belong to the same service, and a \emph{cross-service clone pair} when the two fragments belong to different services~\cite{Mo2021ESEM}.
% サービス内・間クローンの定義を変更
% A clone pair
% is regarded as a within-service clone when both fragments belong to the
% same service, and as a cross-service clone when the fragments belong to
% different services. 
In subsequent work, Zhao et al. analyzed cross-service
clones based on file categories and showed that the distribution and
underlying causes of clones differ across categories~\cite{Zhao2022ICPC}.
% 削減
% Co-modifications of cross-service clones are important from a software
% evolution perspective because they suggest that changes to a cloned
% fragment in one service may require consistent changes to its cloned
% fragments in other services. Such cross-service change coupling can
% undermine service independence and make maintenance more
% difficult~\cite{Mo2021ESEM,Zhao2022ICPC}.
%
% Existing clone detection and visualization tools
% \cite{Ueda2002Gemini,Asaduzzaman2011VisCad,Voinea2014SolidSDD}
% support the inspection of clone detection results, but they are not
% designed for microservice-specific analysis. In particular, they lack
% two capabilities: (i) service-boundary awareness, so within-service and
% cross-service clones are not distinguished and the scope of change
% impact is unclear; and (ii) co-modification awareness, so historical
% change coupling is not integrated, hindering the identification of
% clones that affect maintainability.

These findings indicate that cross-service clones exist, are co-modified, and concentrate in specific file categories, suggesting that such clones may propagate changes across service boundaries and undermine service independence. However, existing clone detection and visualization tools~\cite{Ueda2002Gemini,Asaduzzaman2011VisCad,Voinea2014SolidSDD} organize clone information by file system or program structure (e.g., directories, files, and classes) rather than by service boundaries. In particular, they lack two capabilities: (i) service-boundary awareness, so within-service and cross-service clone pairs are not distinguished and the scope of change impact is unclear; and (ii) co-modification awareness, so historical change coupling is not integrated, hindering the identification of clones that affect maintainability.

To address these limitations, we introduce \textsc{CC4M},
a tool for detecting and
visualizing code clones in microservice systems. It detects Type-2
clones with CCFinderSW, a code clone detector that supports multiple
programming languages~\cite{Semura2017CCFinderSW}, and identifies service
boundaries with CLAIM~\cite{Maggi2024CLAIM}. 
\textsc{CC4M} then enriches each clone pair with service boundary, co-modification, and file-category information, the last of which is inspired by the file-level analysis of Zhao et al. \cite{Zhao2022ICPC}.
% 3つのビューを紹介
Based on the enriched information, \textsc{CC4M} computes clone metrics,
such as service span and co-modification
frequency. The enriched clone pairs are visualized through three views:
a \emph{Statistics View} that summarizes project-wide clone statistics,
an interactive \emph{Scatter Plot} with explicit service boundaries,
and a \emph{Metric View} for metric-based prioritization and
drill-down across services, clone sets, and files.

The contributions of this paper are as follows.
First, we present \textsc{CC4M}, a microservice-aware clone analysis tool
that enriches clone pairs with service-boundary, co-modification,
and file-category information.
Second, we introduce clone metrics, such as service span and
co-modification frequency, to quantify the impact scope and change
coupling across service boundaries.
Third, we provide a scatter plot with explicit service boundaries and
a \emph{Metric View} for metric-based prioritization and drill-down.

\section{CC4M}
\label{sec:CC4M}

\begin{figure*}[tb]
\centering
\includegraphics[width=1.0\linewidth]{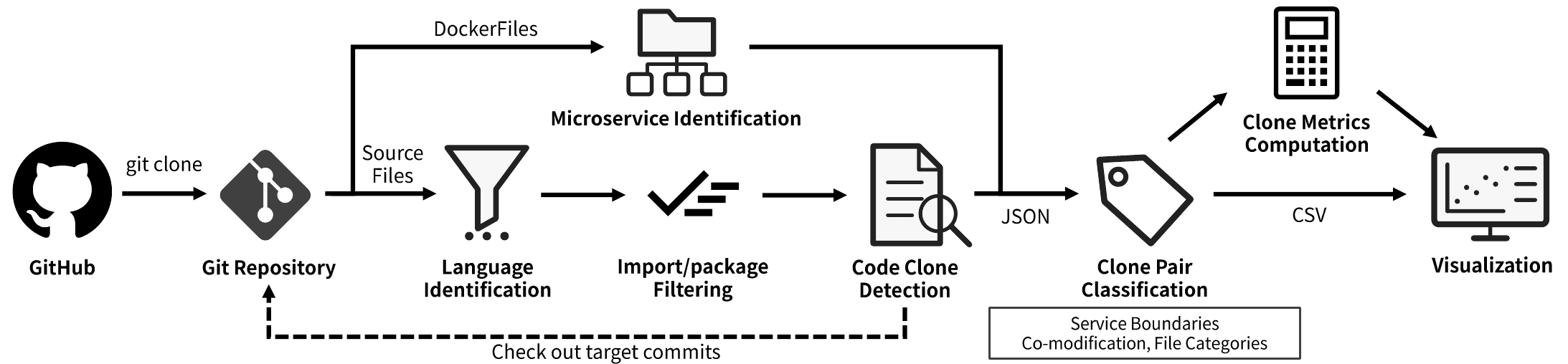}
\caption{
An overview of \textsc{CC4M}. It identifies microservices from a Git repository, detects code clones,
enriches them with service boundaries, co-modification history, and file categories, computes clone metrics, and visualizes the results.
}
\label{fig:overview}
\end{figure*}

Fig.~\ref{fig:overview} shows an overview of \textsc{CC4M}. It takes as
input the GitHub URL of a Dockerized microservice system and provides
three views for exploring code clones: a \emph{Statistics View}, a \emph{Scatter
Plot}, and a \emph{Metric View}. To support these views, \textsc{CC4M} identifies
microservice boundaries, detects Type-2 clone pairs, and enriches each
pair with service-boundary, co-modification, and file-category information.
% CC4Mも斜体
% Hereafter,(MS-Base)

% Figure~\ref{fig:overview} は，CC4M の概要を示している．
% CC4M の入力は，Docker 化されたマイクロサービスシステムの GitHub URL であり,
% コードクローンを分析するための 3 つのビュー,
% すなわち Statistics View, Scatter Plot, および Metric View を提供する.
% これらのビューを支えるため, CC4M は, マイクロサービス境界を識別し,
% Type-2 クローンペアを検出し, 各ペアにサービス境界, 同時修正履歴, および
% ファイルカテゴリ情報を付与する.

\subsection{Microservice Identification}
\label{sec:service_identification}

\textsc{CC4M} identifies microservices by applying CLAIM~\cite{Maggi2024CLAIM},
a lightweight static analysis approach based on Docker Compose files
and related Dockerfiles. CLAIM extracts service definitions from the
Docker configuration files and estimates which directories correspond to application microservices.

\textsc{CC4M} then assigns each source file to a microservice based on the
directory information obtained by CLAIM. 
This assignment classifies each clone pair as within-service or cross-service.

% CC4M は，Docker Compose ファイルおよび関連する Dockerfile に基づく
% 軽量な静的解析手法である CLAIM~\cite{Maggi2024CLAIM} を用いて，
% マイクロサービスを識別する．CLAIM は Docker の構成ファイルから
% サービス定義を抽出し，アプリケーションのマイクロサービスに対応する
% ディレクトリを推定する．
%
% その後，CC4M は CLAIM によって得られたディレクトリ情報に基づいて，
% 各ソースファイルをマイクロサービスへ対応付ける．この対応関係は，
% 各クローンペアがサービス内クローンであるか，サービス間クローンであるかを
% 判定するために用いられる．

\subsection{Code Clone Detection}
\label{sec:clone_detection}

\textsc{CC4M} identifies the programming languages used in the target
repository using GitHub Linguist\footnote{\url{https://github.com/github-linguist/linguist}}
and selects source files whose languages are supported by CCFinderSW.

Before running CCFinderSW, \textsc{CC4M} comments out import, include,
and package declarations in a temporary analysis copy to exclude them
from the token sequence used for clone detection and reduce trivial clones.
The full list of filtered patterns is documented in our GitHub repository.\footnote{\url{https://github.com/gg117e/CC4M/blob/main/docs/declaration_filtering.md}}

\textsc{CC4M} then uses CCFinderSW~\cite{Semura2017CCFinderSW} to obtain
Type-2 clone pairs. Type-2 clones are clone pairs whose fragments are
identical except for differences in whitespace, comments, and identifiers
such as variable names and type names~\cite{RoyCordy2009}.
%CCFinderSW supports multiple programming languages through ANTLR grammar
%definitions, which makes it suitable for heterogeneous microservice
%repositories.

\textsc{CC4M} converts the detection results into a common JSON
intermediate format and passes them to the subsequent analyses. Owing
to this loosely coupled design, the detector is a replaceable
component: supporting another detector requires only a converter to
this format.
%\textsc{CC4M} converts the detection results into a common JSON
%intermediate format using
%ccfindersw-parser\footnote{\url{https://github.com/YukiOhta0519/ccfindersw-parser}}
%and passes them to the subsequent analyses. Owing to this loosely
%coupled design, the detector is a replaceable component.
The current
implementation adopts CCFinderSW as the default detector because
(i) its ANTLR-based grammar definitions cover the multiple languages of
polyglot microservice repositories, (ii) it requires no training data
and behaves deterministically, and (iii) it is lightweight enough for
the detection repeated at every analyzed commit. Recent
deep-learning-based detectors handle Type-3/4 clones well, but
reportedly generalize poorly beyond their mostly single-language (Java)
training benchmarks~\cite{Choi2023ICPC}.

\subsection{Co-modification Analysis}
\label{sec:comodification}

\textsc{CC4M} identifies clone pairs that were modified together based on
Git history.
Users can select commits at a specified interval, merge commits, or
commits associated with Git tags, depending on the desired analysis
granularity.

For each selected target commit, \textsc{CC4M} performs clone detection and
tracks code fragments across adjacent target commits using line
correspondences obtained from \texttt{git diff}.
% クローン
Note that this tracking follows each fragment individually based on the
line correspondences; it does not persist clone-pair relations across
commits. 
For a tracked fragment,
\textsc{CC4M} regards it as modified at commit $c_i$ when lines in its
corresponding range are added, deleted, or changed between $c_{i-1}$
and $c_i$.

% コミット集合Cの定義をする
Let $A$ and $B$ be the two fragments forming a clone pair, and let
$C_A$ and $C_B$ be the sets of target commits in which $A$ and $B$ were
modified, respectively. \textsc{CC4M} regards the clone pair as co-modified
when $C_A \cap C_B \neq \emptyset$, that is, when both fragments were
modified in the same commit at least once.

% CC4M は，Git の変更履歴に基づいて，同時に修正されたクローンペアを特定する．
% 利用者は，分析粒度に応じて，一定間隔のコミット，マージコミット，
% または Git タグに対応するコミットを選択できる．

% CC4M は，各対象コミットにおいてコードクローン検出を実行し，
% \texttt{git diff} から得られる行対応関係を用いて，
% 隣接する対象コミット間でコード片を追跡する．
% なお, この追跡はクローンペアという関係を永続化するものではなく,
% 各フラグメントを git diff の行対応に基づき個別に追跡する.
% 追跡されたコード片について，$C_{i-1}$ から $C_i$ への変更において，
% 対応する範囲内の行が追加，削除，または変更された場合，
% CC4M はそのコード片がコミット $C_i$ において修正されたとみなす．

% クローンペアを構成する 2 つのコード片を $A$ および $B$ とし，
% それぞれが修正された対象コミットの集合を $C_A$ および $C_B$ とする．
% CC4M は，$C_A \cap C_B \neq \emptyset$ である場合，
% すなわち，両方のコード片が少なくとも一度，同一コミット内で修正された場合，
% そのクローンペアを同時修正されたものとみなす．

\subsection{Clone Pair Classification}
\label{sec:classification}

\textsc{CC4M} annotates each clone pair with three kinds of information:
service-boundary, co-modification, and file-category.
Using the service mapping computed in
Section~\ref{sec:service_identification}, it classifies each clone pair as either within-service or cross-service.
Using the
co-modification information computed in Section~\ref{sec:comodification},
it records whether the pair has been co-modified in the Git history.

It also classifies each file into one of four categories: \emph{Test},
\emph{Config}, \emph{Data}, or \emph{Logic}. This classification is based on path and
file-name patterns, and rules are applied in this priority order.
It then assigns a category to each clone pair from the categories
of the two files containing the cloned fragments. Clone pairs whose
two files share the same category are categorized as \emph{Test}, \emph{Config}, or
\emph{Data}; pairs involving both test and non-test files are categorized as
\emph{Mixed}; and the remaining pairs are categorized as \emph{Logic}. The full
classification rules are documented in our GitHub repository\footnote{\url{https://github.com/gg117e/CC4M/blob/main/docs/file_classification.md}}.

% 分類は，パスおよびファイル名のパターンに基づいて行う．
% 概要として，パスまたはファイル名がテストコードを示す場合は
% \textit{Test}，設定ファイルである場合，または設定に関連するディレクトリに
% 存在する場合は \textit{Config}，パスまたはファイル名がエンティティ，
% DTO，データスキーマを示す場合は \textit{Data}，
% それ以外の場合は \textit{Logic} として分類する．
% 複数の規則が 1 つのファイルに一致する場合，CC4M は
% \textit{Test}，\textit{Config}，\textit{Data}，
% \textit{Logic} の順に，最初に一致したカテゴリを割り当てる．
% クローンペアを構成する 2 つのファイルのカテゴリに基づいて，
% CC4M はクローンペアカテゴリを割り当てる．
% 両方のファイルが同じカテゴリに属する場合，そのペアは
% \textit{Test}，\textit{Config}，または \textit{Data}
% として分類される．
% 一方のファイルがテストファイルであり，もう一方が非テストカテゴリに
% 属する場合，そのペアは \textit{Mixed} として分類される．
% それ以外の場合，そのペアは \textit{Logic} として分類される．
% 完全な分類規則は GitHub リポジトリに記載されている\footnote{\url{https://github.com/gg117e/CC4M/blob/main/docs/file_classification.md}}．

\subsection{Clone Metrics Computation}
\label{sec:clone_metrics}

The enriched clone pairs from Sections~\ref{sec:service_identification} to \ref{sec:classification} are numerous, making individual inspection impractical.
To enable metric-based prioritization and filtering, \textsc{CC4M} characterizes clone sets, services, and files with clone metrics.
% \ref{sec:service_identification} 節から \ref{sec:classification} 節までで拡張された
% クローンペアは数が多く, 個別に確認するのは非現実的である.
% メトリクスに基づく優先順位付けとフィルタリングを可能にするため, CC4M は
% クローンセット, サービス, ファイルをクローンメトリクスで特徴付ける.
% 

\textbf{\(\mathbf{MS}(s)\). }
This metric represents the set of microservices to which the code fragments in a given clone set \(s\) belong.
Let \(f \in s\) denote a code fragment in clone set \(s\), and let \(\mathrm{ms}(f)\) denote the microservice to which fragment \(f\) belongs. Then, \(\mathrm{MS}(s) = \{\, \mathrm{ms}(f) \mid f \in s \,\}\), and \(|\mathrm{MS}(s)|\) represents the number of microservices spanned by clone set \(s\).
A clone set \(s\) satisfying \(|\mathrm{MS}(s)| \geq 2\) is called a \emph{cross-service clone set}.

\textbf{\(\mathbf{CM}(s)\).}
This metric represents the set of co-modification commits for a given clone set \(s\).
A commit \(c\) is called a \emph{co-modification commit} of \(s\) if at least two fragments in \(s\) are modified in \(c\).
Then, \(\mathrm{CM}(s)\) is defined as:
\[
\mathrm{CM}(s) = \{\, c \mid \text{at least two fragments in } s \text{ are modified in } c \,\}.
\]
This extends the pair-level co-modification in Section~\ref{sec:comodification}: \(c \in \mathrm{CM}(s)\) exactly when at least one clone pair within \(s\) is co-modified at \(c\).
\(|\mathrm{CM}(s)|\) represents the number of co-modification commits of \(s\).
The co-modification fragment ratio of \(s\), denoted by \(\mathrm{cmr}(s)\), is defined as the fraction of fragments in \(s\) that are modified in at least one commit in \(\mathrm{CM}(s)\).

In the \emph{Scatter Plot} and \emph{Metric View} (see Section~\ref{sec:visualization}), \(|\mathrm{MS}(s)|\), \(|\mathrm{CM}(s)|\), and \(\mathrm{cmr}(s)\) are shown as \textit{service span}, \textit{co-modification frequency}, and \textit{co-modification fragment ratio}, respectively.
In addition to these clone-set-level metrics, \textsc{CC4M} computes per-service and per-file metrics to support filtering and inspection, such as the number of cross-service clone sets, the ratio of cloned lines to service size, and the number of services shared through clone sets.
The full definitions of these additional metrics are documented in our GitHub repository\footnote{\url{https://github.com/gg117e/CC4M/blob/main/docs/metrics.md}}.

\subsection{Visualization}
\label{sec:visualization}

\begin{figure}[tb]
\centering
\includegraphics[width=1.0\linewidth]{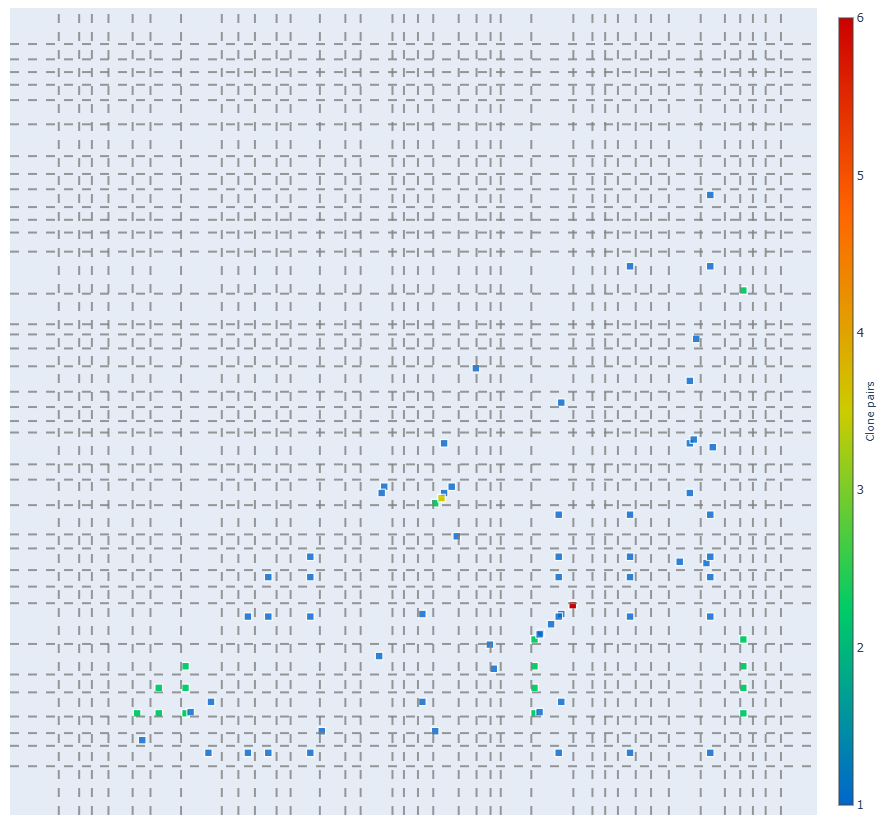}
\caption{
A screenshot of the \textsc{CC4M} scatter plot generated for the \textsc{Train Ticket}
microservice project, with the cross-service filter applied and the
co-modification count set to at least 1. The same source files are arranged on
both axes and grouped by microservice, with gray dotted lines
indicating service boundaries. Each point represents clones between a
pair of files; its color intensity encodes the number of overlapping
clone pairs, and the dark-red point corresponds to the file pair
inspected in Scenario 1.
}
\label{fig:scatter_view}
%(旧キャプション) The Service Scope filter is set to Cross, so the plot shows
%only cross-service clone pairs. Each point represents clones between a
%pair of files, and its color encodes the number of clone pairs.
\end{figure}

\textsc{CC4M} provides three views: (i) a \emph{Statistics View} that reports
clone statistics for the entire project, (ii) a \emph{Scatter Plot} that visualizes
the spatial distribution of clone pairs across service boundaries,
and (iii) a \emph{Metric View} that supports metric-driven filtering
and drill-down across services, clone sets, and files.

\smallskip\noindent\textbf{Statistics View.}
It provides an overview of clones across the project. It summarizes the number of microservices, files, clone sets, and lines of code, breaks them down by language, and ranks the services containing
the most clones.

\smallskip\noindent\textbf{Scatter Plot.}
Fig.~\ref{fig:scatter_view} shows a scatter plot in which the same set
of source files is arranged on both axes in the same order and grouped
by service. Gray dotted lines indicate service boundaries. A clone pair
is plotted at the coordinates of the two files containing the cloned
fragments; color intensity encodes the number of overlapping clone
pairs for the same file pair. Points inside diagonal service blocks represent
within-service clone pairs, whereas points outside represent cross-service
clone pairs. This layout helps users locate clone concentrations across
services without inspecting individual file paths.
A filter panel lets users restrict the displayed clone pairs by service
scope, file category, and co-modification count.
%In
%Fig.~\ref{fig:scatter_view}, Service Scope is set to Cross, so the plot
%shows only cross-service clone pairs. Filters can also be combined; for
%example, restricting file-category to Logic and Co-modification Count to
%at least 1 surfaces cross-service Logic clones that have required
%coordinated changes in the past.
In Fig.~\ref{fig:scatter_view}, the cross-service filter is combined
with a co-modification count of at least 1 to show only cross-service
clone pairs that have required coordinated changes in the past. Users can zoom or pan to inspect dense
regions, and selecting a point opens a detail panel that shows the
corresponding cloned fragments with surrounding source code.
% Fig.に統一
% Code cloneはフラグメントのこと， Code fragment，

\smallskip\noindent\textbf{Metric View.}
It helps users identify services, clone sets, and files that are likely
to have larger maintenance impact according to the computed metrics.
%It supports metric-based prioritization and drill-down
%analysis of clone information. While the \emph{Scatter Plot} helps users
%visually locate clone distributions across service boundaries, the
%\emph{Metric View} helps users identify services, clone sets, and files that
%are likely to have larger maintenance impact according to the computed
%metrics.
% Merric View
The \emph{Metric View} provides three entry perspectives: Microservice-Based
(MS-Based), Clone-Set-Based, and File-Based. In each perspective, users
can narrow down the result list by combining multiple metric filters.
In Clone-Set-Based, users can focus on clone sets that span multiple
services or have past co-modifications by using metrics such as service
span and co-modification frequency. In MS-Based, users can identify
services that contain many cross-service clone sets or have a high
cloned-line ratio. In File-Based, users can find files that share clones
with many services or have cross-service co-modification histories.
The filtered results are shown as ranked lists; selecting a clone set
displays its fragments and their corresponding source code. This staged
workflow, from metric-based prioritization to code-level inspection,
helps users focus on clone relationships that are more likely to affect
maintenance tasks.
%The filtered results are shown as ranked lists, allowing users to
%prioritize items before inspecting source code. When a clone set is
%selected, the \emph{Metric View} displays the fragments contained in the
%clone set. Users can then select one or two fragments to view the
%corresponding source code.

\section{Usage Scenario}
\label{sec:scenario-impact}

This section demonstrates how \textsc{CC4M} supports software evolution
tasks using \textsc{Train Ticket}~\cite{Zhou2018Benchmarking,Zhou2021TrainTicket},
an open-source microservice system selected from a curated benchmark
dataset of Dockerized OSS microservice projects~\cite{Amoroso2024MSR}.
We selected \textsc{Train Ticket} as the subject of the scenarios because
it has been adopted as a benchmark in prior microservice
research~\cite{Mo2021ESEM,Zhao2022ICPC,Cerny2025SANER}.
% 先行研究で同時修正のクローンが発生していることが確認されているという記述をなくして，引用を増やした．
The system is built mainly in Java with a JavaScript frontend.
We use the version of \textsc{Train Ticket} at merge commit
\texttt{813dd01}\footnote{\url{https://github.com/FudanSELab/train-ticket/tree/813dd01}};
all source-code references in this section refer to this version.
For the scenarios in this section, we configure \textsc{CC4M} to detect
co-modification at merge-commit granularity. Clone detection uses
CCFinderSW with a minimum matching token length of 50 and
import/package declaration filtering enabled.
%For the scenarios in this section, we configure \textsc{CC4M} to detect
%co-modification at merge-commit granularity; that is, clone fragments
%are regarded as co-modified when they are modified in the same merge
%commit.
In this snapshot, \textsc{CC4M} identifies 41 microservices, 1{,}423
source files, and 10{,}333 clone sets, most of which are in Java and
JavaScript code. The entire analysis, including clone detection with
CCFinderSW, took 43 minutes on a desktop machine with an Intel Core
i5-14400F CPU.
%In this snapshot, \textsc{CC4M} identifies 41 microservices, 1{,}423 source
%files, 10{,}333 clone sets;
%Table~\ref{tab:train_ticket} reports the per-language breakdown.
%ここはコメントアウト候補
%The following three scenarios illustrate \textsc{CC4M} from different
%perspectives: Scenario 1 focuses on a duplicated booking workflow,
%Scenario 2 on broadly shared data clones with partial co-modification
%follow-up, and Scenario 3 on a repeatedly co-modified Account clone
%that may require further inspection.
We also applied \textsc{CC4M} to six additional microservice systems
selected from the microservice benchmark dataset~\cite{Amoroso2024MSR}.
Table~\ref{tab:projects} summarizes the seven analyzed systems; further
details are available in our project
documentation.\footnote{\url{https://github.com/gg117e/CC4M/blob/main/docs/dataset.md}}
%Further details of these systems are available in our project
%documentation.

\begin{table}[tb]
\centering
\caption{Summary of the analyzed microservice systems.}
\label{tab:projects}
% TODO(gen): 6 システムの Services / Files / LOC / Clone Sets の実測値を記入する.
\begin{tabular}{lrrrr}
\toprule
Project & Services & Files & LOC & Clone Sets \\
\midrule
\textsc{Train Ticket} & 41 & 1{,}423 & 287{,}745 & 10{,}333 \\
FightPandemics & 5 & 436 & 50{,}820 & 755 \\
Lakeside Mutual & 9 & 324 & 37{,}309 & 3{,}431 \\
OpenTelemetry Examples & 28 & 75 & 34{,}962 & 1{,}343 \\
FTGO Application & 10 & 348 & 14{,}308 & 215 \\
IBM Wave7 Lifeline & 10 & 207 & 8{,}914 & 169 \\
Redis Microservices Demo & 9 & 51 & 3{,}919 & 65 \\
\bottomrule
\end{tabular}
\end{table}

% 日本語メモ: 1 つ目のシナリオは, Scatter Plot で cross-service + co-modification
% のフィルタを適用し, 最も濃い点として現れる予約完了ワークフローの並行実装クローンを見つける流れ.
\smallskip\noindent\textbf{Scenario 1.}
Consider a developer who reviews the system to understand where
business logic duplication concentrates across service boundaries.
Without a specific target in mind, they start from the \emph{Scatter Plot},
which gives a visual overview of which service pairs share significant logic
duplication. To narrow the view to such duplication, they apply the
% concentrated logic duplication
cross-service filter and the co-modification filter (Fig.~\ref{fig:scatter_view}).
%In the \emph{Scatter Plot} view, a dark-red point indicates a file
%pair with many overlapping clone pairs; such points are useful starting
%points because they suggest concentrated logic duplication between two
%files.
In the filtered plot, the dark-red point indicates a file pair with
many overlapping clone pairs, suggesting concentrated duplication
between the two files. By clicking this point and selecting the clone set with the most lines in the file pair, they identify the largest cross-service clone pair between the two files.
The two fragments are located in
\href{\repo ts-preserve-service/src/main/java/preserve/service/PreserveServiceImpl.java\#L178-L286}{\texttt{PreserveServiceImpl.java}}
of \mbox{ts-preserve-service} and
\href{\repo ts-preserve-other-service/src/main/java/preserveOther/service/PreserveOtherServiceImpl.java\#L183-L293}{\texttt{PreserveOtherServiceImpl.java}}
of \mbox{ts-preserve-other-service}.
%\texttt{PreserveServiceImpl.java}\footnote{\src{ts-preserve-service/src/main/java/preserve/service/PreserveServiceImpl.java\#L178-L286}{ts-preserve-service/src/main/java/preserve/service/PreserveServiceImpl.java:178-286}}
%of \mbox{ts-preserve-service} and
%\texttt{PreserveOtherServiceImpl.java}\footnote{\src{ts-preserve-other-service/src/main/java/preserveOther/service/PreserveOtherServiceImpl.java\#L183-L293}{ts-preserve-other-service/src/main/java/preserveOther/service/PreserveOtherServiceImpl.java:183-293}}
%of \mbox{ts-preserve-other-service}.
The clone set duplicates the latter part of the booking workflow, including
food order, consign, and notification processing. The \emph{Metric View} shows
service span $|\mathrm{MS}(s)| = 2$, co-modification frequency
$|\mathrm{CM}(s)| = 1$, and co-modification fragment ratio
$\mathrm{cmr}(s) = 1.0$, meaning that both fragments were modified
together in the same commit. This result suggests that future changes
to the booking workflow may need to be applied in both services to
avoid behavioral divergence.

\smallskip\noindent\textbf{Scenario 2.}
Consider a developer tasked with inspecting data definitions that may
cause maintenance problems across service boundaries. In microservice
systems, the same data entity can be duplicated in multiple services,
and inconsistent updates to such copies may lead to subtle bugs.
To find broadly shared data clones with incomplete follow-up, they
filter clone sets by the Data file-category, apply the co-modification
condition, and sort the results by service span in descending order and
by co-modification fragment ratio in ascending order in the
\emph{Metric View}.
This prioritizes data clone sets that are widely distributed but have
incomplete co-modification follow-up. The top entry under these
conditions is a clone set in
which the same \texttt{Order.java} entity is duplicated across nine
services; we cite the copies in
\mbox{\href{\repo ts-order-service/src/main/java/order/entity/Order.java\#L57-L73}{ts-order-service}}
and
\mbox{\href{\repo ts-preserve-service/src/main/java/preserve/entity/Order.java\#L50-L64}{ts-preserve-service}}
as representatives.
%\mbox{ts-order-service}\footnote{\src{ts-order-service/src/main/java/order/entity/Order.java\#L57-L73}{ts-order-service/src/main/java/order/entity/Order.java:57-73}}
%and
%\mbox{ts-preserve-service}\footnote{\src{ts-preserve-service/src/main/java/preserve/entity/Order.java\#L50-L64}{ts-preserve-service/src/main/java/preserve/entity/Order.java:50-64}}
%as representatives.
The clone-set metrics show service span $|\mathrm{MS}(s)| = 9$,
co-modification frequency $|\mathrm{CM}(s)| = 1$, and
co-modification fragment ratio $\mathrm{cmr}(s) = 0.67$. This indicates
that only six of the nine fragments were modified in the past
co-modification commit, leaving three services that did not follow the
update.
%\textsc{CC4M} therefore makes partial follow-up explicit and helps them
%focus inspection on potentially inconsistent data definitions.

\smallskip\noindent\textbf{Scenario 3.}
Consider a developer who assesses the maintenance risk of the system.
They look for cross-service clone sets that have repeatedly required
coordinated changes, since these indicate a recurring maintenance
burden.
To surface them, they filter clone sets with a co-modification frequency
of two or more in the \emph{Metric View}. This filtering identifies a clone set
in which the same \texttt{Account.java} entity is duplicated across
three booking-related services:
\mbox{\href{\repo ts-preserve-service/src/main/java/preserve/entity/Account.java\#L1-L38}{ts-preserve-service}},
\mbox{\href{\repo ts-cancel-service/src/main/java/cancel/entity/Account.java\#L1-L38}{ts-cancel-service}},
and
\mbox{\href{\repo ts-preserve-other-service/src/main/java/preserveOther/entity/Account.java\#L1-L38}{ts-preserve-other-service}}.
%\mbox{ts-preserve-service}\footnote{\src{ts-preserve-service/src/main/java/preserve/entity/Account.java\#L1-L38}{ts-preserve-service/src/main/java/preserve/entity/Account.java:1-38}},
%\mbox{ts-cancel-service}\footnote{\src{ts-cancel-service/src/main/java/cancel/entity/Account.java\#L1-L38}{ts-cancel-service/src/main/java/cancel/entity/Account.java:1-38}},
%and
%\mbox{ts-preserve-other-service}\footnote{\src{ts-preserve-other-service/src/main/java/preserveOther/entity/Account.java\#L1-L38}{ts-preserve-other-service/src/main/java/preserveOther/entity/Account.java:1-38}}.
The clone-set metrics show service span $|\mathrm{MS}(s)| = 3$,
co-modification frequency $|\mathrm{CM}(s)| = 2$, and
co-modification fragment ratio $\mathrm{cmr}(s) = 1.0$. These metrics
indicate that all three fragments were modified together twice.
This repeated co-modification suggests that future changes to the
Account entity may again require coordinated updates across the three
services.

\section{Related Work}

Mo et al. \cite{Mo2021ESEM} reported that code clones exist within
and across microservices in open-source projects, and that some
clones are co-modified in the same version. Zhao et al.
\cite{Zhao2022ICPC} subsequently analyzed cross-service clones in 22
Java microservice projects and reported that 56.7\% of cross-service
clone pairs occur in data-related files. These findings indicate that 
co-modification and file category are useful
perspectives for characterizing cross-service clones, and \textsc{CC4M}
treats both as attributes of clone pairs.

%\begin{table}[tb]
%\centering
%\caption{Comparison with existing tools. SB: service-boundary
%awareness, CM: co-modification awareness, MF: metric-based filtering,
%CV: clone visualization.}
%\label{tab:comparison}
%\begin{tabular}{lcccc}
%\toprule
%Tool & SB & CM & MF & CV \\
%\midrule
%Gemini~\cite{Ueda2002Gemini} & $\times$ & $\times$ & $\times$ & \checkmark \\
%VisCad~\cite{Asaduzzaman2011VisCad}, SolidSDD~\cite{Voinea2014SolidSDD} & $\times$ & $\times$ & \checkmark & \checkmark \\
%DENIM~\cite{Andre2025DENIM}, MicroKarta~\cite{Manglaras2024FSE} & \checkmark & $\times$ & -- & $\times$ \\
%\textsc{CC4M} & \checkmark & \checkmark & \checkmark & \checkmark \\
%\bottomrule
%\end{tabular}
%\end{table}

For visualization, Baker introduced the dotplot to display duplicated
code \cite{Baker1992Duplicated}, and Gemini \cite{Ueda2002Gemini}, the
visualization frontend of CCFinder~\cite{CCFinder}, places files on the two axes of a
scatter plot to provide an overview of clone distributions. VisCad
\cite{Asaduzzaman2011VisCad}, SolidSDD \cite{Voinea2014SolidSDD}, and
the work of Choi et al. \cite{Choi2011IWSC} support metric-based
filtering and refactoring-candidate extraction.
%from clone detection
%results.
%However, these existing approaches do not explicitly account for
%microservice boundaries or co-modification history across services.
Recent microservice visualization tools such as DENIM
\cite{Andre2025DENIM} and MicroKarta \cite{Manglaras2024FSE} visualize
data access points or architectural structure, but do not target clone
relationships.
%to support software
%evolution
No existing tool integrates
service boundaries and co-modification history into clone
visualization, so no directly comparable baseline exists for a
quantitative comparison; the accuracy of the underlying components is
reported in the original
papers~\cite{Semura2017CCFinderSW,Maggi2024CLAIM}. \textsc{CC4M} addresses
this gap by overlaying explicit service boundaries on scatter-plot axes
and by defining service-aware clone metrics such as service span and
co-modification frequency.

% --- 圧縮後 日本語版 (Plan B 適用後) ---
% Mo ら \cite{Mo2021ESEM} は, オープンソースプロジェクトのマイクロサービスにおいて,
% サービス内およびサービス間にコードクローンが存在し, その一部が同一コミット内で
% 同時修正されていることを報告した. Zhao ら \cite{Zhao2022ICPC} はさらに 22 個の
% Java マイクロサービスプロジェクトを分析し, サービス間クローンペアの 56.7\% が
% データ関連ファイルに存在することを示した. これらの知見は,
% 同時修正情報とファイルカテゴリがサービス間クローンを理解するための
% 有用な分析軸であることを示しており, CC4M はこれらをクローンペアの属性として
% 付与する.
%
% 可視化に関して, Baker は重複コードを表す dotplot を導入し
% \cite{Baker1992Duplicated}, CCFinder~\cite{CCFinder} の可視化フロントエンドである
% Gemini \cite{Ueda2002Gemini} は散布図の 2 軸にファイルを配置することで
% クローン分布を俯瞰可能にした. VisCad \cite{Asaduzzaman2011VisCad},
% SolidSDD \cite{Voinea2014SolidSDD}, および Choi らの研究 \cite{Choi2011IWSC}
% は, クローンメトリクスを用いた絞り込みやリファクタリング候補の抽出を支援する.
%しかし, これらのクローン可視化ツールはマイクロサービスを想定して設計されておらず,
%サービス内クローンとサービス間クローンを明示的に区別せず,
%サービス境界をまたぐ同時修正履歴も統合しない.
% 一方, DENIM \cite{Andre2025DENIM} や MicroKarta \cite{Manglaras2024FSE}
% のような近年のマイクロサービス可視化ツールは, データアクセス点や
% アーキテクチャ構造を可視化するが, クローン関係を対象としていない.
% サービス境界と同時修正履歴を
% クローン可視化に統合した既存ツールは無く, 定量比較が可能な直接の
% ベースラインも存在しない. 構成要素の精度は原著論文で報告されている
% ~\cite{Semura2017CCFinderSW,Maggi2024CLAIM}.
% CC4M は, 散布図の軸上に明示的なサービス境界を重ね,
% service span や co-modification frequency などサービスを意識した
% クローンメトリクスを定義することでこのギャップを埋める.

\section{Limitations}
\label{sec:limitations}

\textsc{CC4M} relies on Type-2 clone detection with CCFinderSW. Therefore, structurally similar but partially modified clones, known as Type-3 clones, fall outside its detection scope, and such cross-service clones may be missed.
Moreover, the co-modification analysis covers only clone pairs detected
as Type-2 clones in the analyzed snapshot, that is, the version given
to \textsc{CC4M}. Pairs whose fragments diverged beyond Type-2
similarity before the snapshot are not detected there and are therefore
out of scope.

%It also relies on CLAIM for service-boundary identification, so it
%applies only to microservice systems configured with Docker Compose and
%Dockerfiles.
Applying \textsc{CC4M} requires (i) a service configuration described
with Docker Compose and Dockerfiles, which CLAIM assumes, 
(ii) source code in languages supported by CCFinderSW through ANTLR grammar definitions,
and (iii) a Git history for the co-modification analysis.
In addition, because CLAIM estimates service
directories through static analysis, the estimated boundaries may
contain errors. Prior work reports a microservice identification
accuracy of 82.0\%~\cite{Maggi2024CLAIM}. Such errors directly affect
the classification of clone pairs as within-service or cross-service.

Finally, the clone metrics serve as prioritization aids rather than
validated risk measures, because prior work reports that many clones are
rarely changed and that inconsistent changes are infrequent~\cite{Goode2011}.
\textsc{CC4M} therefore helps users focus inspection rather than
automatically flagging clones as defects.

% \smallskip\noindent\textbf{File classification.}
% The file-category classification is a heuristic based on path and
% file-name patterns, so misclassification can occur in projects that do
% not follow common naming conventions. Files that match no rule are
% treated as Logic, so the Logic category may contain heterogeneous files.
% 

% \section{制限事項}
% \label{sec:limitations}

% \textsc{CC4M} は Type-2 クローンペアを取得するために CCFinderSW を使用しているため，
% 文の挿入，削除，または変更を含む Type-3 クローンを検出することはできない．
% その結果，構造的には類似しているが部分的に変更されたサービス間クローンは見逃される可能性がある．
% さらに, 同時修正分析の対象は, 分析対象スナップショット, すなわち CC4M に
% 与えたバージョンにおいて Type-2 クローンとして検出されたペアに限られる.
% スナップショット以前に Type-2 の類似度を超えて乖離したペアは検出されず,
% 対象外となる.

%\textsc{CC4M} はサービス境界の識別に CLAIM を利用している．
%そのため，Docker Compose および Dockerfile によって構成されたマイクロサービスシステムを前提としており，
%そのような設定ファイルを使用していないプロジェクトには直接適用できない．
% CC4M の適用には, (i) Docker Compose と Dockerfile によるサービス構成
% (CLAIM の前提), (ii) CCFinderSW の文法定義が対応する言語のソースコード,
% (iii) 同時修正分析のための Git 履歴が必要である.
% さらに，CLAIM は静的解析によってサービスディレクトリを推定するため，
% 推定された境界には誤りが含まれる可能性がある．
% 先行研究では，マイクロサービス識別の精度が 82.0\% であることが報告されている~\cite{Maggi2024CLAIM}．
% このような誤りは，クローンペアをサービス内クローンまたはサービス間クローンとして分類する結果に直接影響する．

% 最後に，クローンメトリクスは優先順位付けを支援するための指標であり，
% 検証済みのリスク指標ではない．
% 先行研究では，多くのクローンはほとんど変更されず，
% 不整合な変更も頻繁には発生しないことが報告されている~\cite{Goode2011}．
% したがって，\textsc{CC4M} はクローンを欠陥として自動的に指摘するのではなく，
% ユーザが調査対象を絞り込むことを支援する．

\section{Conclusion and Future Work}

% 日本語メモ: 結論では, CC4M がマイクロサービス識別, Type-2 クローン検出, 同時修正分析, ファイルカテゴリ分類, クローンメトリクスを統合し, サービス境界付きで可視化することをまとめる. TrainTicket の 3 シナリオで, 複数サービスにまたがる協調的変更が必要になり得るクローン関係 (業務ワークフローの並行実装, 部分追従の Data クローン, 反復同時修正されたエンティティ) を発見できることを示した点を強調する. Future Work は Type-3 対応と開発者評価の 1 文に絞る.

% 4行くらい
This paper presented \textsc{CC4M}, a microservice-aware visualization tool for
analyzing code clones. \textsc{CC4M} integrates microservice identification,
Type-2 clone detection, co-modification analysis, file-category
classification, and clone metrics, and visualizes the enriched clone
data with explicit service boundaries. 
%Through three usage scenarios on
%\textsc{Train Ticket}, we showed that \textsc{CC4M} helps users identify clone
%relationships that may require coordinated changes across services.
Future work will extend \textsc{CC4M} to Type-3 clones by integrating
Type-3-capable detectors such as MSCCD~\cite{MSCCD}, and will evaluate
\textsc{CC4M}'s usefulness with developers performing microservice
maintenance tasks.
%Future work will extend \textsc{CC4M} to Type-3 clones and evaluate \textsc{CC4M}'s
%usefulness with developers performing microservice maintenance tasks.

% 本論文では, マイクロサービスを対象としたクローン可視化ツール CC4M を
% 提案した. CC4M は, マイクロサービス識別, Type-2 クローン検出,
% 同時修正分析, ファイルカテゴリ分類, クローンメトリクスを統合し,
% サービス境界付きでクローン情報を可視化する.
% Train Ticket の 3 シナリオを通じ, 複数サービスにまたがる協調的変更が
% 必要になり得るクローン関係を CC4M が発見できることを示した.
%
% 今後の課題として, MSCCD~\cite{MSCCD} のような Type-3 対応検出器を統合する
% ことで CC4M を Type-3 クローンへ拡張し, マイクロサービス保守作業を行う
% 開発者を対象として有用性を評価する.
%今後の課題として, CC4M を Type-3 クローンに対応させ, マイクロサービス
%保守作業を行う開発者を対象として CC4M の有用性を評価する.

\section*{Acknowledgment}
% 科研費を追加
We are grateful to Prof. Toshihiro Kamiya of Shimane University and
Prof. Eunjong Choi of Kyoto Institute of Technology for their
insightful comments on this work.
This work was supported in part by JSPS KAKENHI Grant Numbers JP24K02923 and JP23K28065.
Portions of the implementation were developed with the assistance of Claude Code.
Claude and ChatGPT were also used for English translation and proofreading.
All content was reviewed and validated by the authors.

\bibliographystyle{IEEEtran}
\bibliography{references}

\end{document}